\documentclass[
 reprint,
superscriptaddress,
preprintnumbers,
 amsmath,amssymb,
 aps,
 prl,
]{revtex4-2}
\usepackage{graphicx}
\usepackage{dcolumn}
\usepackage{bm}
\usepackage{xcolor}
\usepackage{physics}
\usepackage{multirow}
\usepackage{orcidlink}
\usepackage[normalem]{ulem}
\usepackage{color}
\usepackage{amsmath}
\usepackage{amssymb}
\usepackage{nicefrac}
\usepackage{tikz}
\usetikzlibrary{positioning}

\newcommand{\casa}{Center for Advanced Simulation and Analytics (CASA), Forschungszentrum J\"{u}lich, Germany}
\newcommand{\ias}{Institute for Advanced Simulation 4 (IAS-4), Forschungszentrum J{\"u}lich, Germany}
\newcommand{\jsc}{J\"{u}lich Supercomputing Center (JSC), Forschungszentrum J\"{u}lich, Germany}
\newcommand{\hiskp}{Helmholtz Institute for Radiation and Nuclear Physics (HISKP), University of Bonn, Germany}
\newcommand{\regensburg}{Institut für Theoretische Physik, Universit{\"a}t Regensburg, Germany}
\newcommand{\tra}{Transdisciplinary Research Area (TRA) Matter, University of Bonn, Germany}
\newcommand{\uvi}{Department of Chemical and Physical Sciences, College of Science and Mathematics, University of the Virgin Islands, US Virgin Islands}
\newcommand{\old}{Oldendorff Carriers GmbH \& Co.\ KG, Lübeck, Germany}
\begin{document}

\title{Simulating Correlated Electrons with Symmetry-Enforced Normalizing Flows}
\author{Dominic Schuh\orcidlink{0000-0003-0866-1404}}\email{schuh@hiskp.uni-bonn.de}
    \affiliation{\tra}
    \affiliation{\hiskp}
\author{Janik Kreit\orcidlink{0009-0000-2454-2339}}\email{jkreit@uni-bonn.de}
    \affiliation{\tra}
    \affiliation{\hiskp}
\author{Evan Berkowitz\orcidlink{0000-0003-1082-1374}}%
    \affiliation{\jsc}
    \affiliation{\ias}
    \affiliation{\casa}
    \affiliation{\uvi}
\author{Lena Funcke\orcidlink{0000-0001-5022-9506}}
    \affiliation{\tra}
    \affiliation{\hiskp}
\author{Thomas Luu}%
    \affiliation{\hiskp}
    \affiliation{\ias}
\author{Kim A.~Nicoli\orcidlink{0000-0001-5933-1822}}
    \affiliation{\tra}
    \affiliation{\hiskp}
    \affiliation{\old}
\author{Marcel Rodekamp\orcidlink{0000-0002-7313-4750}}
    \affiliation{\regensburg}

\newcommand{\ztwo}{{\mathbb{Z}_2}}
\newcommand{\zfour}{{\mathbb{Z}_4}}
\newcommand{\bftheta}{{\boldsymbol{\theta}}}
\newcommand{\Nx}{\ensuremath{N_x}}
\newcommand{\Nt}{\ensuremath{N_t}}
\newcommand{\Ncfg}{\ensuremath{N_\mathrm{cfg}}}
\newcommand{\Z}{\ensuremath{\mathcal{Z}}}
\newcommand{\im}{\ensuremath{\mathrm{i}}}

\date{\today}

\begin{abstract}
\noindent
We present the first proof of principle that normalizing flows can accurately learn the Boltzmann distribution of the fermionic Hubbard model—a key framework for describing the electronic structure of graphene and related materials. State-of-the-art methods like Hybrid Monte Carlo often suffer from ergodicity issues near the time-continuum limit, leading to biased estimates. Leveraging symmetry-aware architectures as well as independent and identically distributed sampling, our approach resolves these issues and achieves significant speed-ups over traditional methods.

\end{abstract}
\maketitle
\emph{Introduction.}---Understanding systems of strongly correlated electrons is fundamental to both condensed matter systems and quantum mechanics~\cite{Arovas_2022}. One of the most widely used theoretical frameworks for studying such systems is the Hubbard model~\cite{hubbard,hubbard1964electronII,hubbard1964electronIII}, which captures the essential competition between electron kinetic energy and on-site interactions.

Over the years, a variety of methods have been developed to analyze the Hubbard model. 
In the weak interaction regime, perturbative approaches provide valuable insights~\cite{Razmadze2024-mi}.
However, outside this regime, non-perturbative effects become significant, rendering perturbative techniques insufficient. In these regimes, Monte Carlo simulations become an indispensable tool (see, for example,~\cite{Ostmeyer2020-od, Ostmeyer2024-ml, Sinilkov2025-xb, Rodekamp2025-wm,Buividovich:2018yar,Beyl:2017kwp,Smith_2014,Ulybyshev_2013} and references therein).
Monte Carlo techniques, such as Hybrid (or Hamiltonian) Monte Carlo (HMC), can suffer from (practical) ergodicity problems
when potential barriers appear between modes of the sampled Boltzmann distribution~\cite{PhysRevB.100.075141}. These barriers can trap the sampler,
leading to biased estimates of physical observables.
A recent study addressed this issue by enhancing HMC with radial updates~\cite{temmen2024overcomingergodicityproblemshybrid}.

In this work, we develop the first deep generative machine learning approach to the Hubbard model, designed to overcome ergodicity issues and enable efficient sampling from the Boltzmann distribution.
We construct and train a symmetry-aware generative model and systematically compare its accuracy and efficiency to that of traditional HMC methods.

Deep generative models---also known as generative neural samplers (GNSs) or, in some contexts, Boltzmann generators~\cite{doi:10.1126/science.aaw1147}---have demonstrated strong capabilities for modeling Boltzmann distributions of physical and chemical systems. 
In recent years, GNSs have been applied across a range of fields, including 
    lattice quantum field theory~\cite{PhysRevD.100.034515,PhysRevLett.126.032001,Caselle:2022acb,cranmer2023advances},
    statistical mechanics~\cite{PhysRevLett.122.080602,PhysRevE.101.023304},
    string theory~\cite{Caselle:2023uel,Caselle:2023mvh},
    and quantum chemistry~\cite{doi:10.1126/science.aaw1147,gebauer1,gebauer3}.
Prominent examples for GNSs include normalizing flows (NFs)~\cite{rezende2015variational,kobyzev2020normalizing,nfreview}
and autoregressive neural networks~\cite{oord2016pixelrecurrentneuralnetworks,NIPS2016_b1301141}.
These models not only enable efficient sampling, but also facilitate the direct estimation of thermodynamic observables~\cite{PhysRevLett.126.032001}, entanglement entropies~\cite{bialas2024r,Bulgarelli:2024yrz}, and Feynman propagators~\cite{PhysRevD.107.056001}. Additionally, GNSs have been employed to guide advanced sampling techniques, such as Feynman-Hall resampling~\cite{Abbott2025-uy}.

Recent studies have shown that generative neural samplers excel at sampling from probability distributions with complex topologies, including bimodal structures~\citep{Nicoli:2021inv,PhysRevD.108.114501} and gauge theories subject to topological freezing~\cite{PhysRevLett.125.121601}. These findings establish GNSs as a promising alternative to traditional methods such as HMC, particularly in regimes where maintaining ergodicity is challenging. In this work, we leverage a novel GNS architecture, called \textit{Symmetry-Enforcing Stochastic Modulation} (SESaMo) \cite{sesamo}, to further enhance sampling performance. 

The remainder of this paper is structured as follows: we first review the Hubbard model, the framework of NFs, and the SESaMo architecture; we then present our numerical results and conclude with a discussion of our findings.

\emph{The Hubbard Model.}---Condensed matter systems can often be modeled as a fixed spatial lattice of atoms---ignoring vibrations and other degrees of freedom---on which electrons can hop between valence orbitals and interact via their electric charge. 
In this work, we employ the Hubbard model~\cite{hubbard,hubbard,hubbard1964electronII,hubbard1964electronIII,brower2012hybridmontecarlosimulation} to describe the dynamics and interactions of the electrons,
\begin{equation}\label{eqn:hubbard}
    H = - \sum_{\langle x,y \rangle,\sigma} c_{x,\sigma}^\dagger c_{y,\sigma} -\frac{U}{2} \sum_x \left(n_{x,\uparrow} - n_{x,\downarrow} \right)^2\ .
\end{equation}
Here, $c_{x,\sigma}^\dagger$ ($c^{}_{x,\sigma}$) denotes the electron creation (annihilation) operator at site $x$ with spin $\sigma$ (=$\uparrow\downarrow$), and $n_{x,\sigma}=c^\dagger_{x,\sigma}c^{}_{x,\sigma}$ is the corresponding number operator.  The first term on the right-hand side of Eq.~\eqref{eqn:hubbard} describes the \textit{tight-binding} dynamics, while the second term captures the onsite repulsive interaction between electrons via the coupling $U$.  Equation~\eqref{eqn:hubbard} is defined such that the electrically neutral half-filled state corresponds to the global ground state of the system.
For stochastic simulations of this system, it is well known that this model can suffer from ergodicity problems~\cite{PhysRevB.100.075141,temmen2024overcomingergodicityproblemshybrid,Beyl:2017kwp,Buividovich:2018yar}.

This quantum mechanical formulation may be rather intuitive to understand, but it is highly unwieldy to handle in practice.
After all, a direct calculation involves a Hamiltonian matrix whose size grows exponentially with the number of lattice sites $\Nx$. It is therefore common to cast the theory into an auxiliary field formulation~\cite{hubbard,hubbard1964electronII,hubbard1964electronIII,brower2012hybridmontecarlosimulation,luu2016QuantumMonteCarlo,rodekamp-phd}.
This reformulation begins by expressing observables in terms of statistical quantum mechanical expectation values:
\begin{equation}
    \expval{\mathcal{O}} = \frac{1}{\Z} \tr{ e^{-\beta H} \mathcal{O} },
    \label{eq:Hilbert-space expectation value}
\end{equation}
where the partition function is given by $\Z = \tr{ e^{-\beta H} }$, $\beta = \nicefrac{1}{k_B T}$ is the inverse temperature, and the trace runs over the entire Fock-space.

This expectation value can be expressed in terms of an auxiliary bosonic field $\phi$, introduced by a Hubbard-Stratonovich transformation~\cite{Hubbard:1959ub},
\begin{equation}
    \expval{\mathcal{O}} = \frac{1}{\Z} \int \mathcal{D}\left[\phi\right] e^{- S\left[\phi\right]} \mathcal{O}\left[\phi\right],
    \label{eq:auxiliary-field expectation value}
\end{equation}
with the Hubbard action given by 
\begin{align}\label{eq:HM_action}
    S[\phi] = \frac{1}{2\tilde{U}}\sum_{x,t\in\Lambda} \phi_{xt}^2 -
         \log \det \left(M \left[\phi \right] \cdot M [-\phi]\right),
\end{align}
where $\tilde{U}=U\beta/N_t$, $\Lambda=\Nx\times \Nt$ is the space-time volume of the lattice, $xt$ labels a site in this volume, $\phi_{xt}\in \mathbb{R}$ is the corresponding field component, and $M$ is the fermion matrix in exponential discretization~\footnote{
    Other choices for incorporating the hopping term into the fermion matrix include the \textit{diagonal} discretization~\cite{brower2012hybridmontecarlosimulation, Luu_2016} and the \textit{linear} discretization~\cite{Smith_2014}. For more details, we refer to the Supplemental Material~\cite{SupMat}.} (see Eq.~\eqref{eq:fermion_matrix} in the Supplemental Material~\cite{SupMat}).

The Hubbard action consists of two main contributions: a Gaussian term, encoding interactions, and a fermionic term, describing the fermion-dynamics. Importantly, the fermion determinant is manifestly positive~\cite{PhysRevB.100.075141}, which ensures that the Boltzmann weight can be interpreted as a proper probability distribution.

The goal of this work is to provide the first proof of principle that NFs can be applied to the Hubbard model.
In doing so, we demonstrate how ergodicity and autocorrelation issues can be essentially eliminated. We focus on a spatial lattice extent fixed to two sites, i.e., $\Nx=2$, where these effects are particularly pronounced, while we vary the temporal extent $\Nt$. 

An additional benefit of studying the two-site system is that the exact configuration distributions are analytically known in the strong-coupling limit, enabling direct comparisons with our GNS flows in this regime. We also benchmark our results against analytically known correlation functions~\cite{PhysRevB.100.075141}
\begin{equation}\label{eqn:C(t)}
    C_{x,y}(t)\equiv\langle c^{}_x(t)c^\dagger_y(0)\rangle=\frac{1}{\Ncfg}\sum_{\{\phi\}}M^{-1}_{(t,x);(0,y)}[\phi]\ ,
\end{equation}
where $N_{\rm cfg}$ is the number of sampled configurations and $M$ is the fermion matrix (see Eq.~\eqref{eq:fermion_matrix} in \cite{SupMat}). The correlator is an \emph{observable} that provides information about the interacting spectrum of the system.
The correlation functions are computed for all site pairs $(x,y)$, and subsequently diagonalized to obtain eigenfunctions $C_\pm(t)$ \cite{PhysRevB.100.075141}.

\emph{Symmetries of the Hubbard Action.}---The symmetries of this system have been well documented (see, e.g., Ref.~\cite{PhysRevB.100.075141}).
Of particular interest here are the symmetries of the fermion matrix $M$ and its determinant.
Reference~\cite{PhysRevB.100.075141} showed that, as a consequence of sub-lattice (also referred to as \textit{chiral}) symmetry and particle-hole symmetry, one obtains
\begin{align*}
    \det (M[\phi]M[-\phi]) &= f[\phi]f[-\phi]\\
    f[\phi]=e^{-\Phi/2}\det M[\phi]&=e^{\Phi/2}\det M[-\phi]=f[-\phi]\ ,
\end{align*}
where $\Phi=\phi_1+\phi_2$, with $\phi_i=\sum_t\phi_{i,t}$.
The function $f$ is therefore an even function of the field variables.  
For the two-site model, this has important consequences in the strong coupling limit ($U\gg1)$, where the determinant becomes nearly invariant under $\phi_i\to-\phi_i$.  
As a result, the system exhibits an approximate $\zfour$ symmetry in the variables $\phi_1$ and $\phi_2$.
As we will show later, explicitly incorporating this symmetry into our neural network architecture is crucial for correctly capturing the probability distribution and observables, and for preventing mode collapse.

\emph{Normalizing Flows.---}Normalizing flows \cite{kobyzev2020normalizing,nfreview} are generative models that enable efficient sampling from high-dimensional distributions by learning an invertible map between a simple base distribution $q_z$ (e.g., a Gaussian) and a complex target distribution $q_y$ defined by the physical system. Here, $z$ denotes latent variables, and $y$ corresponds to the physical degrees of freedom—in our case, the auxiliary Hubbard field configurations $\phi$. This map is realized through a bijective transformation
\begin{equation}
    f_\theta: z \sim q_z \mapsto y \sim q_y \,,
\end{equation}
where $f_\theta$ is a bijective function parameterized by neural network weights $\theta$.
The primary objective is to construct a bijective map $f_\theta$ that is sufficiently expressive to model the target distribution. 
In recent years, numerous types of NFs have been proposed, including coupling-based NFs \cite{nice,realnvp,splineflows}, autoregressive NFs \cite{autoregressive}, and continuous NFs \cite{continuousflow}.
In this work, we focus on coupling-based NFs with a real-valued non-volume preserving (RealNVP) architecture~\cite{realnvp}. These consist of $L$ invertible and differentiable transformations $f^l_{\theta_l}: \mathbb{R}^{|\Lambda|} \to \mathbb{R}^{|\Lambda|}$, where each transformation is parametrized by a neural network with parameters $\theta_l$. The bijectivity of each individual transformation ensures that the overall map is also bijective
\begin{equation}
    y \equiv f_\theta(z) = \left( f^L_{\theta_L} \circ f^{L-1}_{\theta_{L-1}} \circ \dots \circ f^1_{\theta_1} \right) (z) \,,
\end{equation}
where $\theta = \{ \theta_1, \hdots, \theta_L \}$ collectively parametrizes the map $f_\theta$. In order to enable density estimation of $q_y$, it is essential to efficiently compute the Jacobian determinant of $f_\theta$ during training.

The incorporation of physics priors, such as symmetries, has been shown to significantly improve the performance and training stability of generative models~\cite{groupequivariant2016,sphericalcnn2018,Boyda_2021}. In  Ref.~\cite{sesamo}, the \textit{Symmetry-Enforcing Stochastic Modulation} (SESaMo) framework was introduced, which enhances NFs by incorporating stochastic modulation. This approach enforces symmetries on the output density by randomly applying symmetry transformations during sampling. In the context of this work, we employ a stochastic modulation
\begin{equation}
    S_\zfour \equiv S_2 \circ S_1: y \sim q_y \mapsto \phi \sim q_\bftheta\, ,
\end{equation}
which imposes an approximate $\zfour$ symmetry to $q_\bftheta$ by sequentially applying two stochastic transformations: a $\ztwo$-symmetric modulation $S_1$, and an \textit{approximate} $\ztwo$-symmetric modulation $ S_2$.

The first transformation, $S_1$, flips the sign of the entire field $y$ with a probability of $50\,\%$
\begin{align}
    S_1: y \mapsto \begin{cases}
        \;\;\,y &\text{if } u_1 = 0 \\
        -y &\text{if } u_1 = 1
    \end{cases} && \text{where} && u_1 \sim \mathcal{B}(0.5) \,.
\end{align}
The second transformation, $S_2$, selectively flips the components $y_{1}=(y_{11},y_{12},\dots,y_{1N_t})^T$ with a probability controlled by a \textit{learnable} breaking parameter $b \in \mathbb{R}^-$
\begin{align}
     S_2: y_{1} \mapsto \begin{cases}
        \;\;\,y_{1} &\text{if } u_2 = 0 \\
        -y_{1} &\text{if } u_2 = 1
    \end{cases} && \text{where} && u_2 \sim \mathcal{B}(e^b) \,,
\end{align}
with the Bernoulli distribution $\mathcal{B}(\cdot)$. Thus, $S_1$ enforces exact $\ztwo$ symmetry, while $S_2$ introduces a controlled symmetry \textit{breaking} in the components $y_{1t}$, with the degree of breaking learned during training via the parameter $b$.

The complete transformation used to generate the final field configuration $\phi$ is thus given by
\begin{equation}
    \phi \equiv g_\bftheta(z) = \left( S_2 \circ S_1 \circ f_\theta \right) (z) \,,
\end{equation}
where $\bftheta = \{ \theta, b \}$ denotes the complete set of trainable parameters of the model. The resulting output probability density is expressed by
\begin{equation}
    q_\bftheta(g_\bftheta(z)) = q_z(z) \cdot \left| \frac{\partial f_\theta}{\partial z} \right|^{-1} p_{S_1} \cdot p_{S_2}(u_2) \,,
\end{equation}
where $p_{S_1}$ and $p_{S_2}$ are the probabilities associated with the stochastic modulation steps $S_1$ and $S_2$, respectively, which are given by
\begin{align}
    p_{S_1} = \frac{1}{2} && \text{and} && p_{S_2}(u_2) = \begin{cases}
        1 - e^b &\text{if } u_2 = 0 \\
        \;\;\;e^b &\text{if } u_2 = 1 \,.
    \end{cases}
\end{align}
The objective of the training process is to minimize the \textit{self-regularized} KL divergence, an extension of the reverse KL divergence \cite{kullback1951information}, which is used to train the NF with SESaMo. This divergence measures the distance between the model output density $q_\bftheta$ and the target Boltzmann density $p = e^{-S}/Z$, and is given by
\begin{align}
    \label{eq:self_reg_kl}
    \widetilde{\mathrm{KL}}(q_\bftheta \, || \, p) = \;\mathbb{E}_{z \sim q_z} \bigg[ 
    & \log q_\bftheta[g_\bftheta(z)] + S[g_\bftheta(z)] \nonumber \\
    & + \gamma \log \hat Z + \Lambda(f_\theta(z)) 
    \bigg] \,,
\end{align}
where $S[\cdot]$ is the physical action, and $\hat Z$ is an estimator of the partition function scaled by a hyperparameter $\gamma$. The penalty term $\Lambda$ acts as a regularization term that numerically enforces bijectivity of $g_\bftheta$. Further details are given in the Supplemental Material~\cite{SupMat}. To ensure exactness, we apply an accept/reject step to the generated configurations \cite{PhysRevD.100.034515}. 

\emph{Results.}---In the following, we evaluate the performance of our NF-based approach for the Hubbard model on a two-site lattice with parameters $U=18$ and $\beta=1$. We employ a RealNVP network with an approximate $\zfour$ symmetry enforced via the SESaMo approach. Specifically, we compare the resulting probability densities, physical observables, scaling behavior, and sampling efficiency.

\begin{figure}[b]
    \centering
    \includegraphics[width=0.49\linewidth]{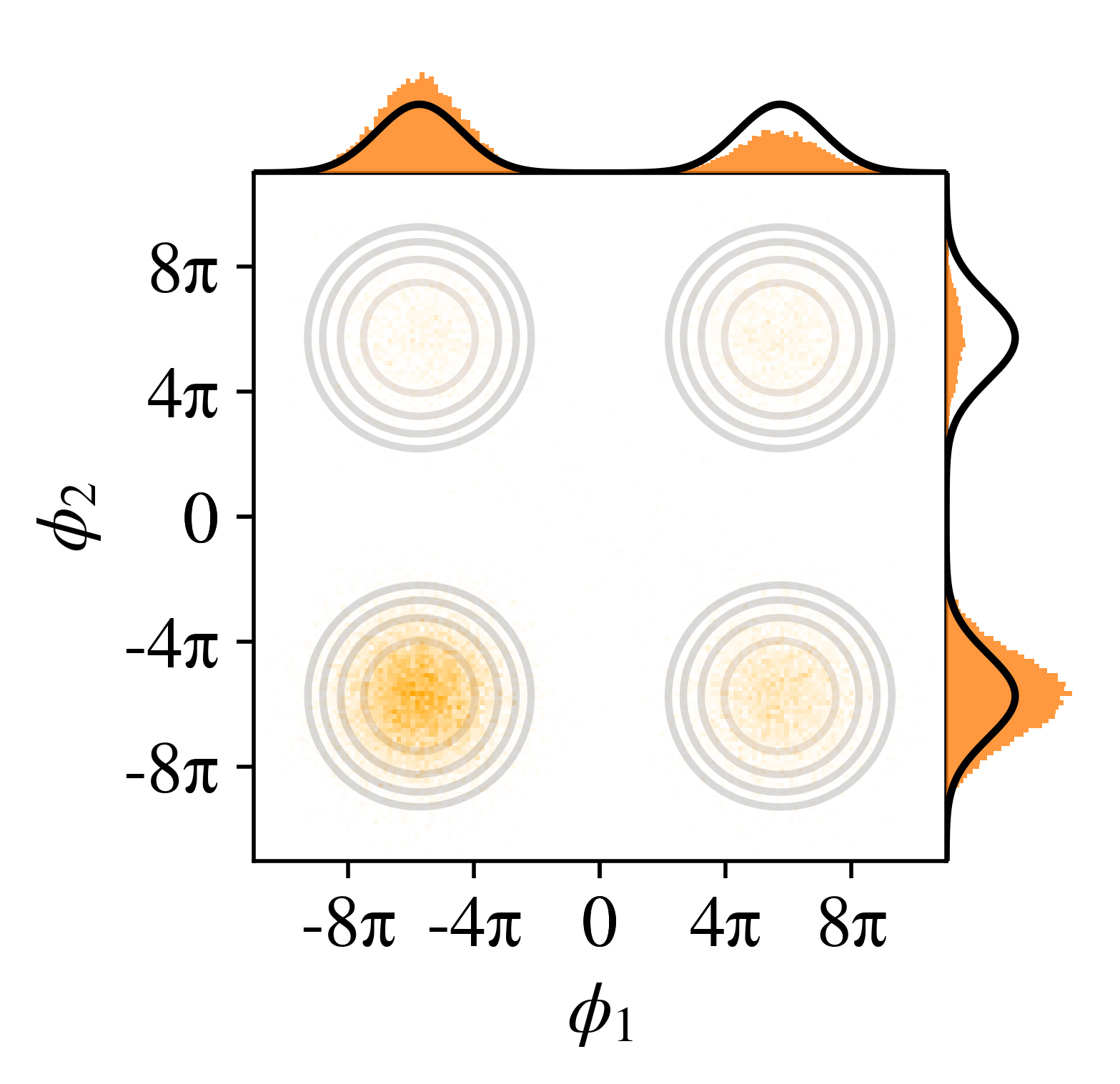}
    \includegraphics[width=0.49\linewidth]{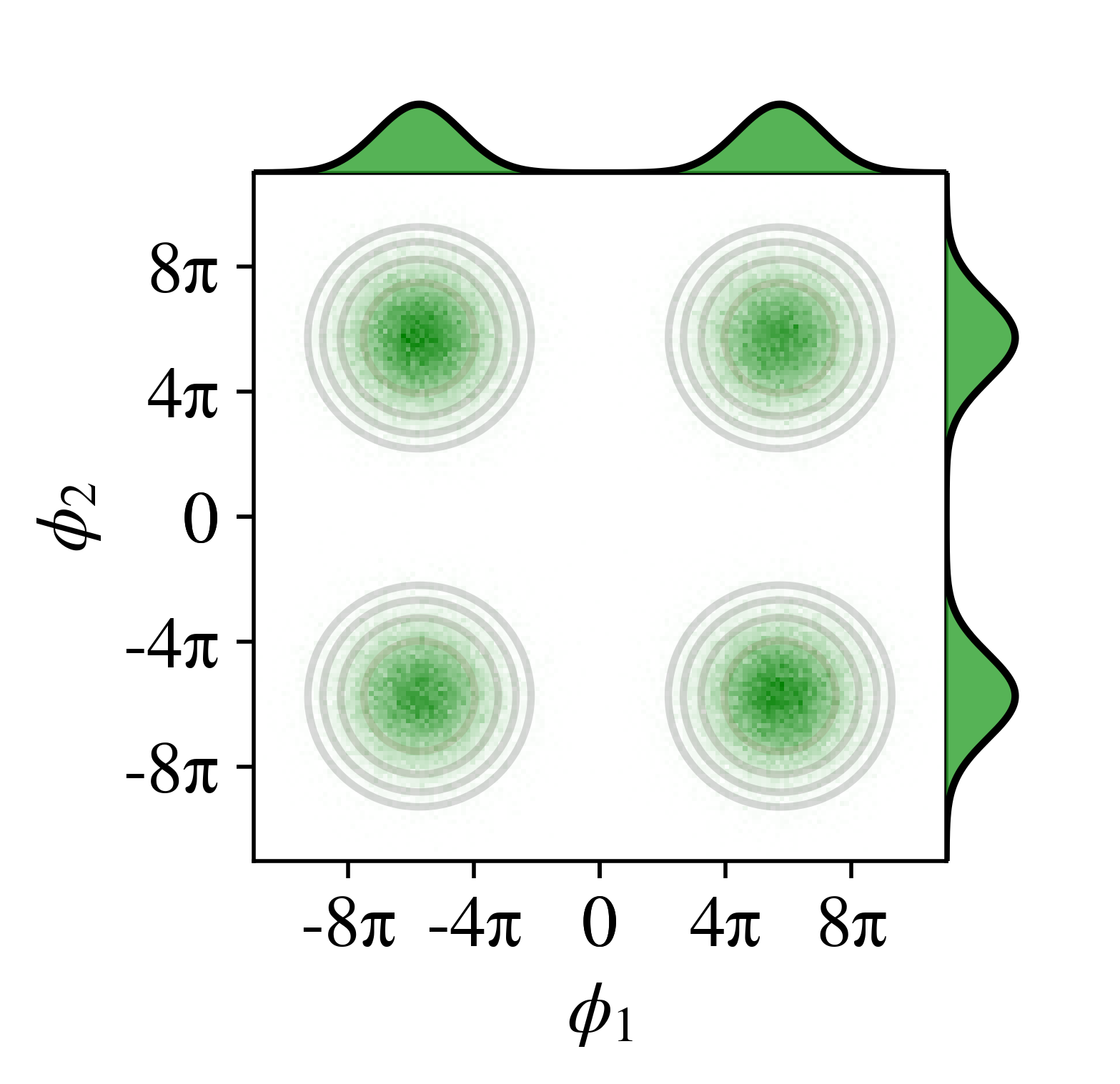}
    \caption{
    Two-site Hubbard model ($U=18$, $\beta=1$): Sampled field configurations from 25k decorrelated samples generated by HMC (orange) and NF (green), summed over the time direction, i.e., $\phi_x = \sum_{t=1}^{\Nt} \phi_{xt}$.
    Gray contours indicate the reference distribution in the strong-coupling limit. The outer histograms show the marginal distributions in $\phi_1$ and $\phi_2$, with the black lines indicating the exact analytical results obtained from Eq.~(70) in Ref.~\cite{PhysRevB.100.075141}.
    }
    \label{fig:2_site_densities}
\end{figure}

In Fig.~\ref{fig:2_site_densities}, we show the sampled distribution obtained from HMC with decorrelation measures (orange, left), i.e., taking only every 10th sample from the Markov chain \cite{PhysRevB.100.075141}, and from the NF (green, right). HMC struggles to distribute the relative weight correctly among all modes, as it predominantly collapses onto the lower-left mode and consequently undersamples the remaining ones, which can result in  biased estimates of physical observables. In contrast, the NF successfully learns the correct multimodal distribution and samples with a high acceptance rate of $a=98.3\,\%$.

\begin{figure}[tb]
    \centering
    \includegraphics[width=0.9\linewidth]{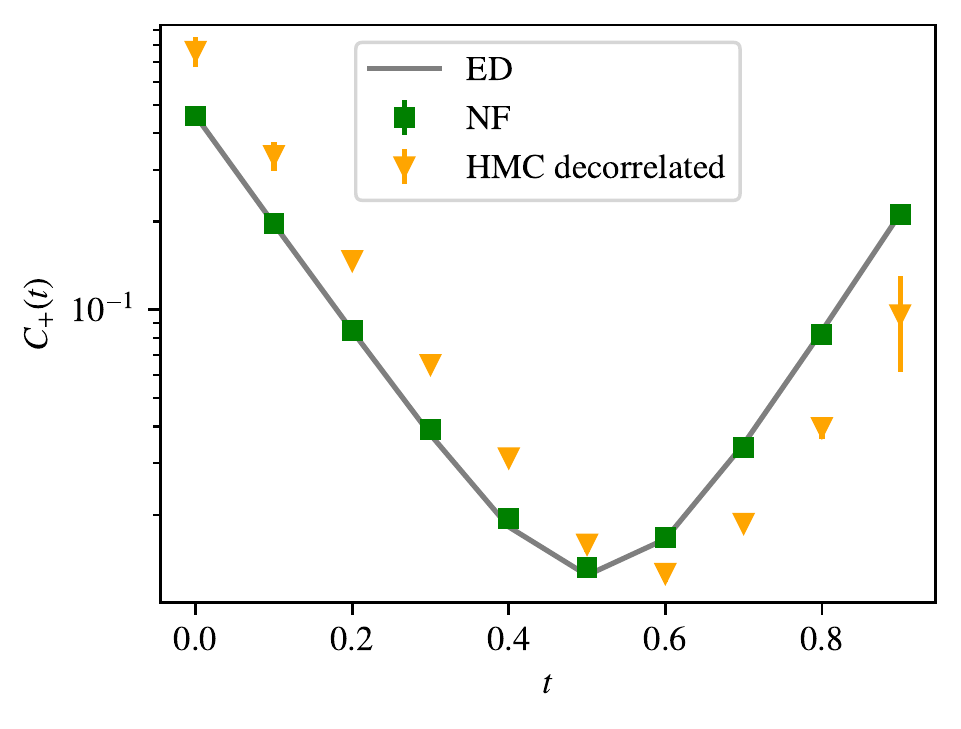}
    \caption{One-body correlation function $C_{+}(t)$ of the two-site Hubbard model ($U = 18$, $\beta = 1$). Results from 25k decorrelated configurations generated by HMC (orange) and the NF (green) are shown, alongside the ground truth obtained from exact diagonalization (ED, gray).
    }
    \label{fig:correlator}
\end{figure}

To further validate the superior performance of the NF, we compute the even one-body correlation function $C_+(t)$ as defined in Eq.~\eqref{eqn:C(t)}, with results shown in Fig.~\ref{fig:correlator}. The figure compares outcomes from HMC with decorrelation measures (orange), NF (green), and exact diagonalization (ED, gray), where ED serves as the ground truth in this scenario. We observe that the strong ergodicity issues in the HMC samples lead to systematically biased results that deviate from the ground truth. In contrast, the NF approach accurately reproduces the observables with excellent agreement and significantly smaller uncertainties. 

\begin{figure}[tb]
    \centering
    \includegraphics[width=0.9\linewidth]{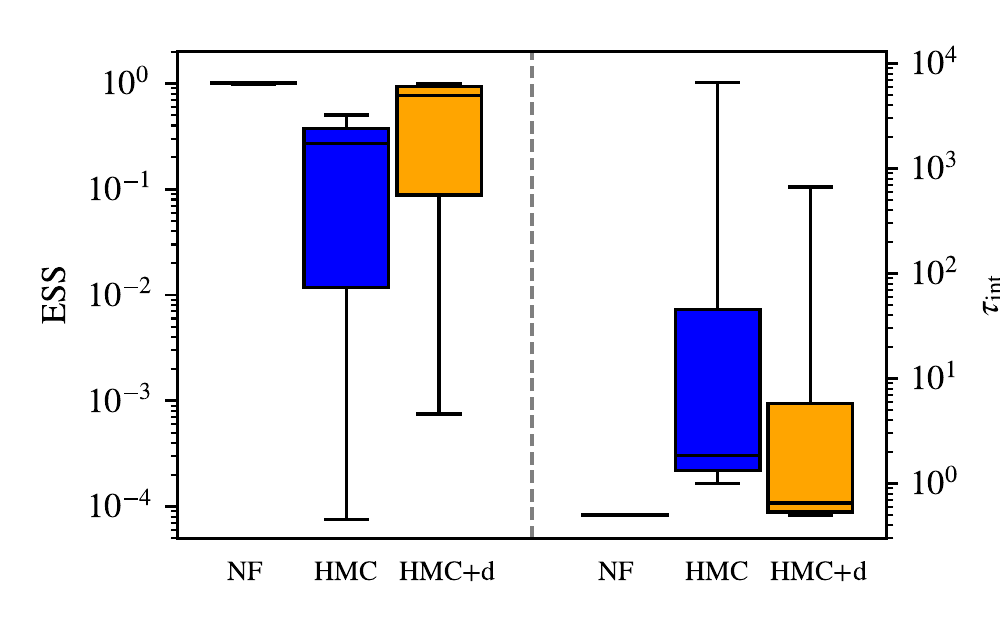}
    \caption{Comparison of the effective sample size (ESS, left axis) and the integrated autocorrelation time $\tau_{\mathrm{int}}$ (right axis) for three sampling methods: NF (green), HMC (blue), and HMC with decorrelation measures (HMC+d, orange). Both metrics are computed from the one-body correlation function shown in Fig.~\ref{fig:correlator}. A higher ESS and lower $\tau_{\mathrm{int}}$ indicate more efficient sampling, underscoring the superior performance of the NF approach.}
    \label{fig:ess_tauint}
\end{figure}

Figure~\ref{fig:ess_tauint} compares the sampling efficiency of the presented methods using the effective sample size (ESS) and the integrated autocorrelation time $\tau_\text{int}$, evaluated on the correlators shown in Fig.~\ref{fig:correlator}. The NF (green) achieves both a higher ESS and a lower $\tau_\text{int}$ compared to HMC without decorrelation measures (blue), demonstrating superior sampling performance. Even when decorrelation measures are applied to HMC  (orange), the NF remains significantly more efficient. These improvements underscore the advantages of (i) generating independent, parallel samples and (ii) incorporating symmetry-aware training through SESaMo.

\begin{figure}[tb]
    \centering
    \includegraphics[width=0.99\linewidth]{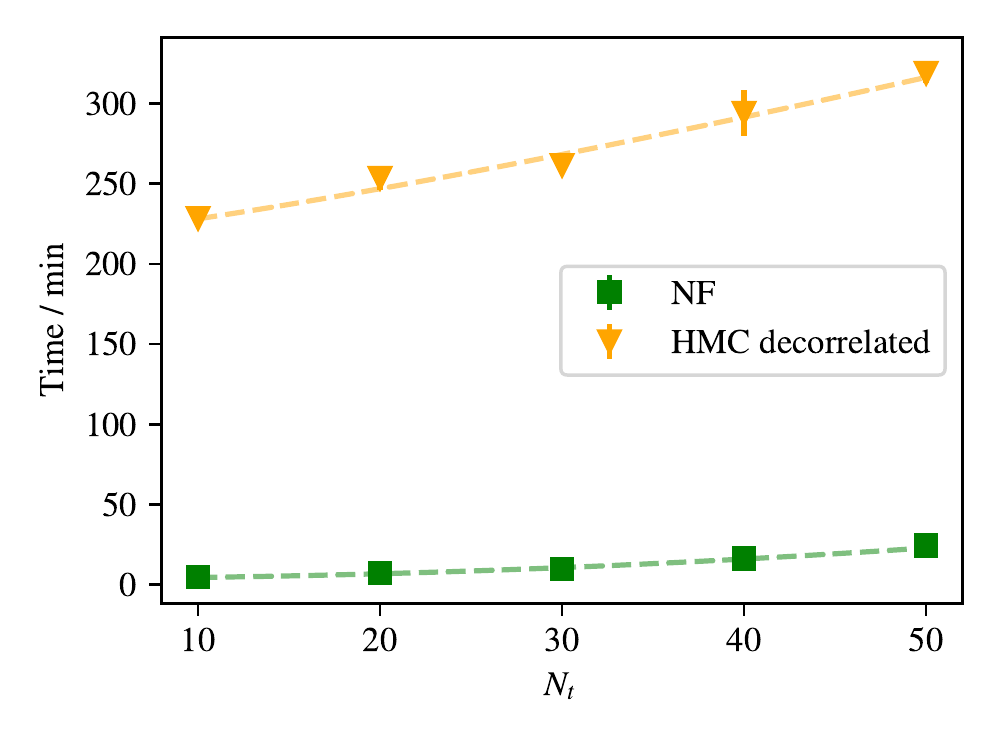}
    \caption{Sampling time---including training time for the NF---as a function of the temporal lattice extent $\Nt$, for generating 25k decorrelated  configurations. HMC is considered thermalized after 10k steps. Results are shown for HMC (orange) and a NF(green). Error bars indicate the statistical uncertainty in the measured time across three independent runs. The dotted lines show fits to the observed scaling behavior: $N_t^{5/4}$ for HMC, consistent with theoretical expectations, and an empirical quadratic dependence for the NF.}
    \label{fig:nt_scaling}
\end{figure}

One of the main limitations of flow-based samplers is their unfavorable scaling with the lattice volume \cite{abbott2023aspects, PhysRevD.104.094507, abbott2023normalizingflowslatticegauge, Komijani:2023Up}, as well as the relatively high computational cost of training. To directly assess the scaling performance of HMC and the NF, we compare the total wall-clock time---\textit{including} training in the case of the NF---required to generate 25k decorrelated configurations across a range of temporal extents $\Nt$, as shown in Fig.~\ref{fig:nt_scaling}. The results show that incorporating symmetry constraints via SESaMo reduces the total sampling time for the NF by up to 98\,\% compared to HMC.

\begin{figure}[tb]
    \centering
    \includegraphics[width=0.9\linewidth]{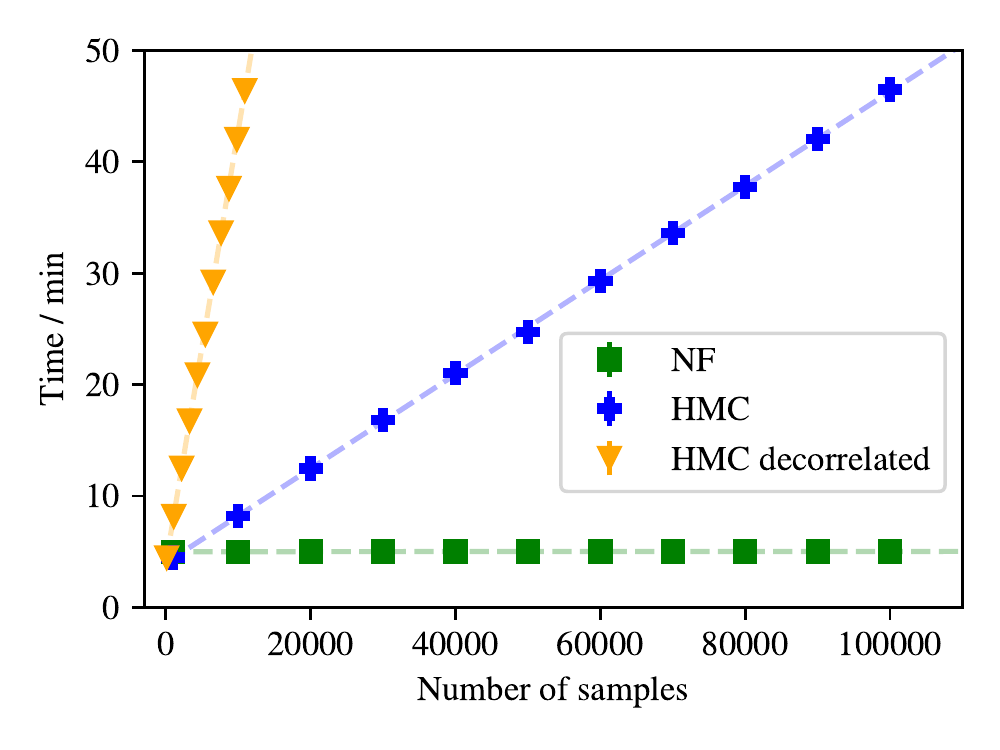}
    \caption{
    Comparison of sampling costs for HMC (blue), HMC with decorrelation measures (orange), and the NF (green), including five minutes of training time for the NF. For HMC, thermalization is assumed after 10k steps. All results correspond to the Hubbard model with $U = 18$, $\beta = 1$, and $\Nt = 10$. The trained flow model achieved an acceptance rate of $97.43\,\%$. The dotted lines indicate a linear fit to the number of samples.}
    \label{fig:sampling_cost_nt10}
\end{figure}

Finally, in Fig.~\ref{fig:sampling_cost_nt10}, we present a comparison of the total sampling time---including training in the case of the NF---for a system with $\Nt = 10$ time slices, using HMC (blue), HMC with decorrelation measures (orange), and the NF (green). We find that, due to the embarrassingly parallel sampling with the NF and the low training overhead enabled by the efficient symmetry incorporation via SESaMo, the total sampling time for the NF increases only marginally beyond an initial training period of approximately 5 minutes. In contrast, the sampling time for HMC grows linearly with the number of configurations and, when applying decorrelation measures, becomes roughly two orders of magnitude larger than that of the NF.

\emph{Conclusion.}---In this work, we presented the first proof of principle that NFs can accurately learn the Boltzmann distribution of the \emph{fermionic} Hubbard model and yield high-precision results for physical observables such as correlators, especially in regimes where HMC fails due to ergodicity issues. Our results establish that NFs can surpass HMC by a wide margin, both in accuracy and computational efficiency. Furthermore, we demonstrated that incorporating symmetry constraints via SESaMo enables favorable (quadratic) scaling with the temporal lattice extent. Additional results for larger spatial lattice sizes are provided in the Supplemental Material~\cite{SupMat}. Taken together, 
%beyond the two-site Hubbard model, 
these findings establish a novel, scalable, and broadly applicable framework for efficient sampling in complex lattice systems, where traditional methods are hindered by ergodicity issues and long autocorrelations.

\emph{Acknowledgments.}---The authors thank Simran Singh, Luca Johannes Wagner, Gurtej Kanwar, and Daniel Hackett for inspiring discussions. This project was supported by the Deutsche Forschungsgemeinschaft (DFG, German Research Foundation) as part of the CRC 1639 NuMeriQS – project no. 511713970.
\bibliography{refs}
\appendix
\section{Basis Choice for the Hubbard Model}
\noindent
In the main part of this letter, we introduced the Hubbard model in the so-called \textit{spin basis}. This choice, however, is not unique. More generally, the Hubbard Hamiltonian can be written in a continuous family of bases as
\begin{align}
H =\,& H_0 + \alpha \,\frac{U}{2} \sum_x (n_{x,\uparrow}-n_{x,\downarrow})^2 \nonumber \\ &- (1-\alpha) \,\frac{U}{2} \sum_x (n_{x, \uparrow} - n_{x, \downarrow})^2,
\end{align}
where the parameter $\alpha \in [0,1]$ interpolates between the spin basis ($\alpha = 0$) and the particle/hole basis ($\alpha = 1$). The two limiting cases correspond to purely real ($\alpha = 0$) and purely imaginary ($\alpha = 1$) Hubbard-Stratonovich fields. Intermediate values of $\alpha$, by contrast, typically lead to complex-valued fields. For this reason, the two limiting bases are especially advantageous for numerical simulations, as they reduce the dimensionality of the field space.

The Hamiltonian in the particle/hole basis thus reads
\begin{equation}
    H_{ph} = -\sum_{\langle x,y\rangle} (a_x^\dagger a_y - b_x^\dagger b_y) + \frac{U}{2} \sum_x (n_x-\tilde{n}_x),
\end{equation}
where $n_x=a^\dagger_x a_x$ and $\tilde{n}_x=b^\dagger_x b_x$ count the number of particles and holes at position $x$, respectively. Following the same steps as in the spin basis, one obtains the Hubbard action in the particle/hole basis
\begin{align}\label{eq:HM_action_ph}
    S_{ph}[\phi] = \frac{1}{2\tilde{U}}\sum_{x,t\in\Lambda} \phi_{xt}^2 -
         \log \det \left(M \left[i \phi \right] \cdot M [-i \phi]\right) \,.
\end{align}
In contrast to the spin basis, the fermion matrix is now evaluated solely on imaginary fields $i\phi$. 

\section{Discretizations of the Fermion Matrix}
\noindent
In the discretized path integral formulation of the Hubbard model, the inverse temperature $\beta$ is evenly divided into $\Nt$ time slices. The resulting fermion matrix $M[\phi]$, which emerges after integrating out the fermionic degrees of freedom, depends on the chosen discretization scheme. While all schemes converge to the same continuum limit $\delta=\beta/N_t\to 0$~\cite{PhysRevB.100.075141}, they differ at finite $\delta$, resulting in distinct matrix structures and ergodicity properties.

In the main analysis of this letter, we focused on the so-called \textit{exponential} discretization,
\begin{align}\label{eq:fermion_matrix}
    M \left[ \phi \right]_{x't',xt} = \delta_{x',x}\delta_{t',t} - [e^h]_{x',x} e^{\phi_{xt}} \mathcal{B}_{t'} \delta_{t',t+1} \, ,
\end{align}
where the hopping matrix is defined as $h_{x',x}= \delta_{\langle x',x\rangle}$, with $\delta_{\langle x',x\rangle}$ restricting to allowed hoppings. The factor 
$\mathcal{B}$ explicitly encodes anti-periodic boundary conditions in the temporal direction $t$. To facilitate comparison with previous studies, we also consider the \textit{diagonal} discretization, in which the exponential of the hopping matrix is expanded to first order. The  resulting fermion matrix takes the form
\begin{equation}
        M^d \left[ \phi \right]_{x't',xt} =( \delta_{x',x}-h_{x',x})\delta_{t',t} -e^{\phi_{xt}} \delta_{x',x} \mathcal{B}_{t'} \delta_{t',t+1} \, .
\end{equation}
Although both discretizations formally agree in the continuum limit~\cite{PhysRevB.100.075141}, they differ in their ergodic properties: the diagonal discretization does not suffer from a formal ergodicity problem, while the exponential discretization in the particle/hole basis encounters codimension-1 manifolds in configuration space (i.e., subspaces of dimension $\Nx\Nt - 1$) on which the probability weight vanishes. As a result, field configurations separated by such manifolds cannot be connected through finite-action paths, and standard Monte Carlo updates may become trapped within disconnected regions of configuration space. These manifolds are depicted as solid black lines in Fig.~\ref{fig:2site_allbases}.

\section{Symmetries of the Hubbard Action in the Particle/Hole Basis}
\label{sec:symmetries_hubbard}
\noindent
The Hubbard model exhibits a rich symmetry structure that depends on the chosen basis. These symmetries can apply to the full action or, more restrictively, only to the fermion determinant. For a comprehensive classification and discussion of the symmetry structure across different formulations, we refer the reader to Ref.~\cite{PhysRevB.100.075141}. Here, we focus on the periodicity symmetry relevant to our analysis in the particle/hole basis.

In the particle/hole basis (i.e., $\alpha = 1$), the probability weight defined in Eq.~\eqref{eq:HM_action_ph} depends only on $e^{i\phi}$. As a result, it is invariant under discrete shifts of the auxiliary field
\begin{align}
\label{eq:periodicity_symmetry}
\phi_{xt} \rightarrow \phi_{xt} + 2\pi\cdot n_{xt}, \quad n_{xt} \in \mathbb{Z} \, .
\end{align}
This transformation leaves the fermion determinant unchanged but does not preserve the Gaussian part of the action. It is therefore only an \textit{approximate} symmetry of the system, valid strictly at the level of the fermionic sector.

\section{Details on the Self-Regularized KL Divergence}
\label{sec:penalty}
\noindent The regularization term in the self-regularized KL divergence, introduced in Eq.~\eqref{eq:self_reg_kl}, depends on the field configuration $y$ after the NF $f_\theta$ and before the stochastic modulation $S$. It is defined as
\begin{equation}
    \label{eq:penalty_term}
    \Lambda(y) = \sum_{x=1}^{\Nx}  A \sigma (B \lambda_x(y)) \cdot \theta(\lambda_x(y)) \,,
\end{equation}
where $\sigma$ and $\theta$ denote the sigmoid and Heaviside functions, respectively. The hyperparameters $A$ and $B$ control the overall strength of the regularization and the steepness of its gradient. The penalty function $\lambda_x(y)$ is given by
\begin{equation}
    \lambda_x(y) = - \sum_{t = 1}^{\Nt} y_{xt} \,.
\end{equation}
Finally, the hyperparameter used to scale the the partition function estimate $\hat{Z}$ is fixed to $\gamma = 0.5$.
\section{Normalizing Flows with Canonicalization}
\label{sec:canonicalization}
\noindent
Similarly to SESaMo, canonicalization can be used to incorporate prior physics knowledge into NFs. The key idea is to construct a map $g_\theta$ that is equivariant under a given symmetry transformation $T$, i.e.,
\begin{equation}
    g_\theta(Tz) = T g_\theta(z)\,.
\end{equation}
In the canonicalization procedure, each sample $z \in \mathbb{R}^{|\Lambda|}$ is mapped into a so-called \textit{canonical cell} $\Omega$ via a canonical transformation $C_{T,z}: z \in \mathbb{R}^{|\Lambda|} \mapsto y \in \Omega$. The NF $f_\theta$ then acts only within this canonical cell, transforming $y \in \Omega$ to $\tilde y \in \tilde \Omega$. Finally, the inverse canonical transformation $C_{T,z}^{-1}: \tilde y \in \tilde \Omega \mapsto \phi \in \mathbb{R}^{|\Lambda|}$ maps the output back to the original field space.

Combining the NF with the canonical transformations, the output field $\phi$ is given by the composition
\begin{equation}
    \phi \equiv g_\theta(z) = \left( C_{T,z}^{-1} \circ f_\theta \circ C_{T,z} \right) (z) \,,
\end{equation}
and the corresponding output density is given by
\begin{equation}
   q_\theta(g_\theta(z)) = q_z(z) \cdot \left| \frac{\partial f_\theta}{\partial z} \right|^{-1} \, .
\end{equation}
As with SESaMo, the flow is trained using the self-regularized KL divergence in Eq.~\eqref{eq:self_reg_kl}. The penalty term $\Lambda(\tilde y)$, defined in Eq.~\eqref{eq:penalty_term}, is applied to restrict the transformed samples $\tilde y = \left( f_\theta \circ C_{T,z} \right) (z)$ to remain within the canonical cell, i.e., $\tilde \Omega \subseteq \Omega$.
For further details and generalizations of this approach, we refer the reader to Ref.~\cite{sesamo}.

\subsection{\texorpdfstring{$2\pi$}{2 pi} Canonicalization}
\label{sec:2pi_canonicalization}
\noindent As discussed previously, the fermion matrix in the particle/hole basis exhibits a $2\pi$ periodicity symmetry. In contrast to previous work \cite{Schuh2025-ef}, it was found to be beneficial to define the canonical cell $\Omega$ such that the field components summed in the time direction,
\begin{equation}
  \phi_x = \sum_{t=1}^{\Nt} \phi_{xt}  \, ,
\end{equation}
are restricted to lie between $-\pi$ and $\pi$. The canonical cell is thus defined as
\begin{equation}
    \Omega = \left\{ y \in \mathbb{R}^{|\Lambda|}: \left| \sum_{t=1}^{\Nt} y_{xt} \right| \leq \pi, \quad \forall \, x = 1, \hdots , \Nx  \right\} \,.
\end{equation}
The canonical transformation
\begin{align}
    C_{2\pi, z}: z_{x0} \mapsto z_{x0} - 2\pi \cdot k_x, 
    && k_x = \text{round} \left( \frac{z_x}{2\pi} \right) \, ,
\end{align}
with $z_x = \sum_{t=1}^{\Nt}z_{xt}$, ensures that all samples $z$ lie inside the canonical cell. The inverse canonical transformation
\begin{align}
    C_{2\pi, z}^{-1}: \tilde y_{x0} \mapsto \tilde y_{x0} + 2\pi \cdot k_x,
    && k_x = \text{round} \left( \frac{z_x}{2\pi} \right) \, ,
\end{align}
maps samples $\tilde y$ back to their original space. Note that $C_{2\pi, z}^{-1}$ depends explicitly on $z$, so information about the original configuration is retained in the transformation.

To enforce that samples remain within the canonical cell, the penalty function
\begin{equation}
    \lambda_x(\tilde y) = \left| \sum_{t=1}^{\Nx} \tilde y_{xt} \right| - \pi \, ,
\end{equation}
is used in the penalty term defined in Eq.~\eqref{eq:penalty_term}.

\section{Additional Results for the Particle/Hole Basis and Diagonal Discretization}
\noindent
To enable direct comparison with the HMC results reported in Ref.~\cite{PhysRevB.100.075141}, we train a NF using the $2\pi$-canonicalization procedure to learn the path integral distribution in the particle/hole basis. Additionally, we extend our analysis to include the diagonal discretization alongside the exponential discretization.
\begin{figure}[t!]
    \centering
    \vspace*{-1.5mm}
    \includegraphics[width=0.9\linewidth]{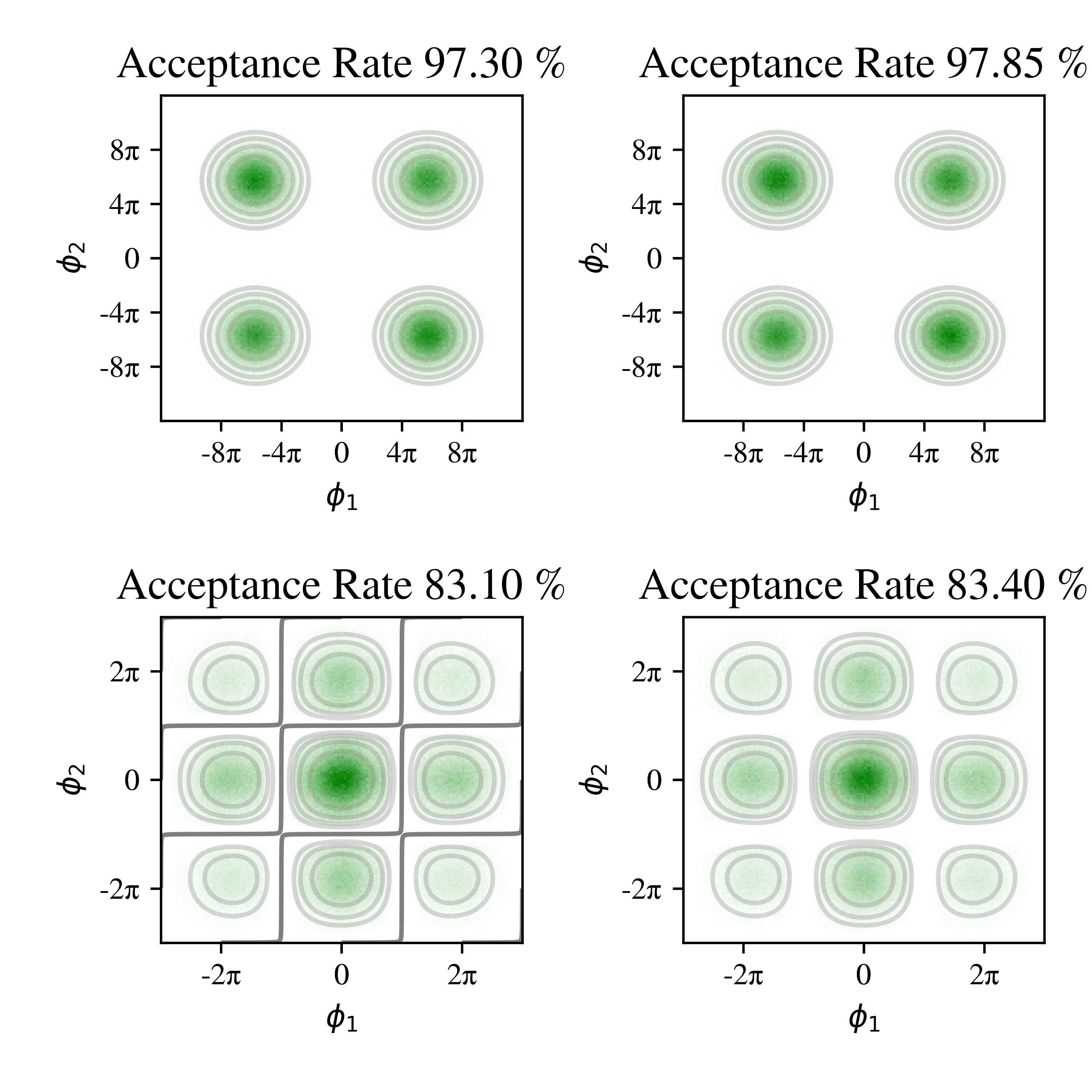}
    \vspace*{-1.5mm}
    \caption{Two-site distributions of the Hubbard model with on-site interaction $U = 18$, inverse temperature $\beta = 1$, and $N_t = 10$ time slices. The top row shows results in the spin basis, and the bottom row in the particle/hole basis. The left column corresponds to exponential discretization, while the right column corresponds to diagonal discretization. Gray contours indicate the underlying probability distribution in the strong-coupling limit. All distributions are generated using a NF: the top row employs $\mathbb{Z}_4$ stochastic modulation, while the bottom row uses $2\pi$ canonicalization. }
    \label{fig:2site_allbases}
\end{figure}
As demonstrated in Fig.~\ref{fig:2site_allbases}, the NF successfully captures the correct structure of the target distribution across all bases and discretizations. 

At comparable acceptance rates, the NF significantly outperforms HMC by generating independent and identically distributed samples with full mode coverage \cite{PhysRevB.100.075141, temmen2024overcomingergodicityproblemshybrid}. 
This advantage is particularly pronounced in the spin basis, where the SESaMo framework enables a more expressive, symmetry-aware representation, further improving accuracy and acceptance rates.

\section{Additional Results for the Four-Site Square Lattice}
\noindent
In the main part of this letter, we focused on the two-site model, where ergodicity issues are particularly pronounced and can be clearly visualized. To demonstrate the scalability and broader applicability of our method, we now present results for a two-dimensional four-site square lattice. At this system size, exact reference data from exact diagonalization (ED) is still available, allowing a direct quantitative comparison.

\begin{figure}[t!]
    \centering
    \includegraphics[width=0.9\linewidth]{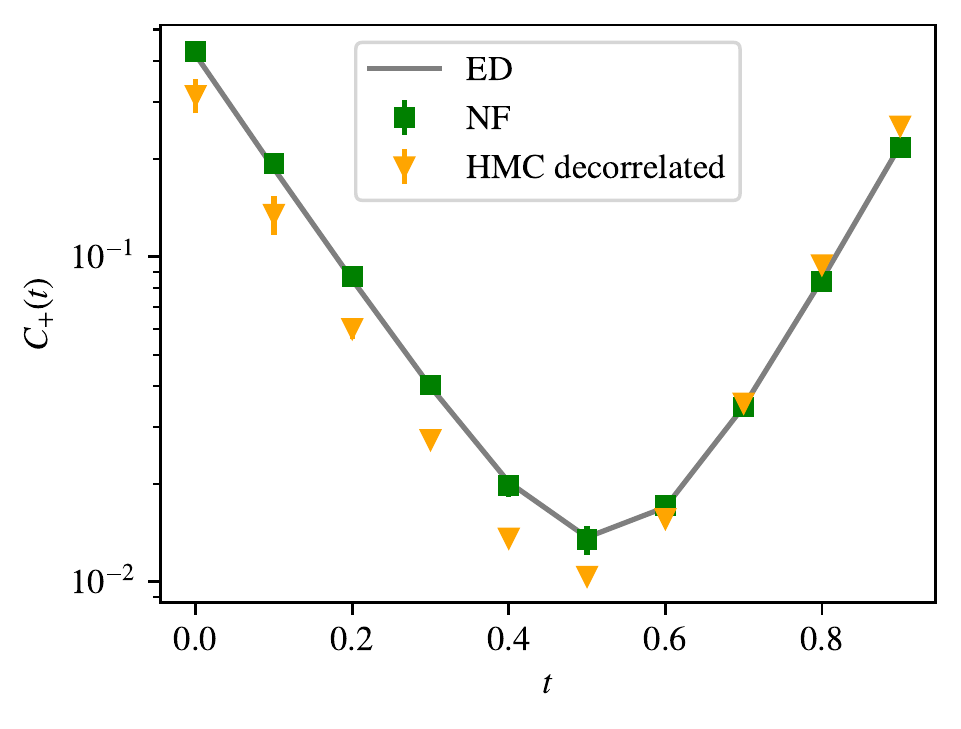}
    \vspace*{-1.5mm}
    \caption{One-body correlation function $C_{+}(t)$ of the four-site Hubbard model on a square lattice with interaction strength $U = 18$ and inverse temperature $\beta = 1$. Results obtained using HMC equipped with decorrelation measures (orange) and a NF (green) are shown, along with the ground truth from exact diagonalization (ED, black). The NF accurately reproduces the ground truth with significantly smaller uncertainties, whereas HMC fails to capture the correct behavior.}
    \label{fig:four_site_correlator}
\end{figure}

We consider a system with interaction strength $U = 18$, inverse temperature $\beta = 1$, and $\Nt = 10$ time slices. As an observable, we examine the one-body correlation function, shown in Fig.~\ref{fig:four_site_correlator}. Similarly to the two-site case, HMC (orange) exhibits clear deviations from the ground truth due to ergodicity issues. In contrast, the NF (green) accurately reproduces the exact result across all time separations.
This result highlights the effectiveness of the symmetry-aware NF beyond the two-site system and demonstrates its applicability to larger lattices. The accurate reproduction of exact observables at this scale suggests that the approach can be reliably extended to even larger system sizes.
\end{document}